\documentstyle[procl]{article}
\def\beq{\begin{equation}}

\def\eeq{\end{equation}}

\def\as{\alpha_s}

\def\bea{\begin{eqnarray}}

\def\eea{\end{eqnarray}}

\begin{document}

\begin{flushright}
\mbox{~}\\[-2cm]
DESY 96-145\\
hep-ph/9608414
\end{flushright}
\vspace{1.5cm}

\title{HEAVY QUARKS PHOTOPRODUCTION$^\heartsuit$}
\footnotetext{$^\heartsuit$Talk given at the DIS'96 Workshop, Rome, 
April '96}

\author{ MATTEO CACCIARI }

\address{Deutsches Elektronen-Synchrotron DESY,\\ D-22603 Hamburg, Germany}

%%%%%%%%%%%%%%%%%%%%%%%%%%%%%%%%%%%%%%%%%%%%%%%%%%%%%%%%%%%%%%

% You may repeat \author \address as often as necessary      %

%%%%%%%%%%%%%%%%%%%%%%%%%%%%%%%%%%%%%%%%%%%%%%%%%%%%%%%%%%%%%%

\maketitle\abstracts{ The state of the art of the theoretical calculations for
heavy quarks photoproduction is reviewed. The full next-to-leading order 
calculation and two possible resummations, the high energy one for total cross
sections and the large $p_T$ one for differential cross sections, 
are described.}

\section{Introduction}

Heavy quarks production processes provide a powerful insight into our
understanding of Quantum Chromodinamics. The large mass of the heavy quark can
make the perturbative calculations reliable, even for total cross sections, by
setting a large scale at which, for instance, the strong coupling can be
evaluated and found small enough. On the experimental side, the possibility to
tag heavy flavoured mesons by means of microvertex detectors can provide quite
accurate measurements.

All these potentialities must of course be matched by an accurate enough 
theoretical evaluation of the production cross section. In this talk I will
describe the state of the art of such calculations for heavy quarks 
photoproduction. I
will firstly review the next-to-leading order (NLO) QCD calculations recently
presented by Frixione, Mangano, Nason and Ridolfi, for both direct and resolved
photons. This calculation, available for total cross sections, one-particle and
two-particle distributions, is now a consolidated result and provides a
benchmark for future developments. Next I will describe the resummation of two
kinds of large logarithms which appear in the  NLO fixed order calculations and
potentially make it less reliable in some regimes: $\log(S/m^2)$ is large when
the center of mass energy $\sqrt{S}$ is much larger than the mass $m$ of 
the heavy quark,
and $\log(p_T^2/m^2)$ becomes large when the transverse  momentum $p_T$ of the
observed quark is much larger than its mass.

\section{Fixed Order Calculation}
Heavy quarks photoproduction at leading order in the strong coupling $\as$
looks a very simple process: only the tree level diagram $\gamma g \to Q\bar Q$
contributes at the partonic level, and the final answer for the total cross
section (to be convoluted with the gluon distribution function) reads
%\beq
$
\hat\sigma_{\gamma g}(\rho) = 2\pi\alpha\as e_Q^2 \rho\beta\left[\left(
1+\rho-{\rho^2\over 2}\right){1\over\beta}\ln{{1+\beta}\over{1-\beta}}
-1-\rho\right]
%\eeq
$
with $\rho=4m^2/S$ and $\beta=\sqrt{1-\rho}$, $m$ being the heavy quark mass
and $\sqrt{S}$ the center-of-mass energy of the interaction. This result is
simple and well behaved, being finite everywhere.

At a deeper thinking, however, problems seem to arise. For instance, one may
ask himself why not to include initial state heavy quarks, coming from the
hadron and to be scattered by the photon, like $\gamma Q \to Q g$. To include
consistently such a diagram is not an easy task, especially if one wants to 
keep the quark heavy. Taking it massless, on the other hand, would not only be
a bad approximation but would also produce a divergent total cross section.
A way out of this was provided by Collins, Soper and Sterman \cite{css}, 
who argued that
for total cross section of heavy quarks in hadron collisions 
the following factorization formula holds:
\beq
\sigma = \sum \int f_{a/H_1} f_{b/H_2} \hat\sigma(ab\to Q\bar Q)
\label{QQfact}
\eeq
The sum on the partons runs only on $a$ and $b$ being gluons or light quarks,
and the heavy quarks are only generated at the perturbative level by gluon
splitting. There is therefore no need to try to accomodate them in the
colliding hadrons and the relevant kinematics can be kept exact. Eq.
\ref{QQfact} provides the basis for an exact perturbative calculation of heavy
quarks production to NLO. For what concerns photoproduction, such a calculation
has been first performed by P.~Nason and K.~Ellis, and subsequently confirmed
by J.~Smith and W.L. van Neerven \cite{en}. When going to order $\alpha\as^2$ 
in photon-hadron collision, however, a new feature appears. The photon can now
couple directly to massless quarks, for instance in processes like $\gamma q
\to Q\bar Q q$, and in a given region of phase space a collinear singularity
will appear. It can be consistently factored out, but this requires the
introduction of a {\sl photon structure function} which, pretty much like
hadron structure functions, will describe the probability that before the
interaction the photon splits into hadronic components (light quarks or gluons,
in this case). Such a behaviour is sometimes called {\sl resolved photon} (as
opposed to {\sl direct}). A full NLO calculation for heavy quark
photoproduction will therefore also require a NLO calculation for
hadroproduction \cite{nde}, where one of the structure funtions will be the 
photon's one.
A factorization scale $\mu_\gamma$, related to the subtraction of the
singularity at the photon vertex, will link the two pieces and its dependence
on the result will only cancel when both are taken into account.

Frixione, Mangano, Nason and Ridolfi \cite{fmnr} have recently presented 
Montecarlo integrators (which are not, we stress, NLO Montecarlo generators) 
for these two calculations, thereby allowing detailed comparisons with
experimental data. Rather then presenting here the plots with these comparisons
I refer you to the original literature \cite{fmnr}. 
The overall result can however be
summarized as follows. Total cross sections seems to be reproduced by the
calculation both at fixed target and HERA regimes, but the huge uncertainties
present both on the experimental and the theoretical side do not allow the
study of finer details like, for instance, the relevance of the resolved
component at HERA. For what concerns transverse momentum distributions at fixed
target, they can be reproduced after allowing for the heavy quark fragmentation
to mesons
and for a primordial transverse momentum of the incoming partons of the order of
1 GeV. These same non-perturbative corrections also allow for a description of
two-particle correlations, thereby pointing towards a consistent picture. On the
other hand, a word of caution is mandatory in the light of experimental 
results presented
at this Workshop \cite{h1-zeus}, which show a pseudorapidity distribution of 
the heavy quark
at HERA not in agreement with the NLO calculation.

\section{High Energy Resummation}
Like any perturbative expansion, the NLO calculation for heavy quarks
photoproduction is only reliable and accurate as long as the coefficients of
the coupling constant remain small. Large terms of the kind $\ln(S/m^2)$ (or
$\ln(1/x)$, since $xS = s \sim m^2$) do however appear in the kernel total
cross section, and for growing $S$ they will
eventually became large enough to spoil the convergence of the series order by
order. Such terms need therefore to be resummed to all orders to allow for a
sensible phenomenological prediction.

Theoretical frameworks for obtaining such a resummation have been provided in
the papers of ref. \cite{small-x}. 
The general procedure is that of replacing the usual
``collinear pole'' factorization for the kernel cross section
\beq
4m^2 \sigma_{\gamma g}(\rho,m^2,Q_0^2) = \int_0^1 
{{dz}\over{z}} \; C\!\left({\rho\over z}, \as(m^2)\right) G(z,m^2,Q_0^2)
\eeq
with  a new
``high-energy'' factorization, where the gluon which couples to the $Q\bar Q$
system is kept off-shell by an amount ${\boldmath k}^2$:
\beq
4m^2 \sigma_{\gamma g}(\rho,m^2,Q_0^2) = \int d^2 k\int_0^1 
{{dz}\over{z}} \hat\sigma\left({\rho\over z}, {{k^2}\over{m^2}}\right) {\cal
F}(z,k,Q_0^2)
\eeq
The corresponding unintegrated gluon structure function ${\cal F}$, 
also depending on
${\boldmath k}$, will obey a BFKL evolution equation whose solution resums
the dangerous small-$x$ terms. 

Not many phenomenological results are available within this framework. Bottom
production at HERA has been estimated to grow by about 10\% when these effects
are taken into account.

\section{Large Transverse Momentum Resummation}

More potentially large terms, of the kind $\ln(p_T^2/m^2)$, do appear when
considering the one particle inclusive differential distribution at large
transverse momentum. These terms should also be resummed to produce a reliable
theoretical prediction in this region. Such a resummation has been performed in
ref. \cite{cg} along the following lines. 
One observes that in the large-$p_T$ limit the
only important mass terms are those appearing in the logs, all the others being
power suppressed. This means that an alternative description of heavy quark
production can be achieved by using {\sl massless} quarks and providing at the
same time structure and fragmentation functions also for the heavy quark:
\beq
\sigma = \sum \int f_{a/H_1} f_{b/H_2} \hat\sigma(ab\to c) d_{Q/c}
\label{fact}
\eeq
Indices $a$,$b$ and $c$ now also run on $Q$, taken massless in $\hat\sigma$.
The key point is that the large mass of the heavy quark allows for 
the evaluation in perturbative QCD of its structure and fragmentation 
functions at a scale given
by its mass. The logs will appear in these function, which  can then be 
evolved with the Altarelli-Parisi 
equations up to the large scale set by $p_T$. This evolution will resum the
large logarithms previously mentioned. Phenomenological analyses show that the
effect becomes sizeable only at very large $p_T$, say greater than 20 GeV for
charm photoproduction, and
should therefore not be very important at HERA.

\vspace{1mm}
\noindent
{\small
{\bf Acknowledgments.} It is a pleasure to thank A. Vogt for inviting me 
to give this talk and the
organizers for the wonderful atmosphere surrounding the whole Conference.
}
\vspace{-8mm}
\section*{References}\vspace{-3mm}
% ----------------------------------------------------------------------------
% ------------------- define commands for some papers -------------------
\newcommand{\zp}[3]{Z.\ Phys.\ {\bf C#1} (19#2) #3}
\newcommand{\pl}[3]{Phys.\ Lett.\ {\bf B#1} (19#2) #3}
\newcommand{\plold}[3]{Phys.\ Lett.\ {\bf #1B} (19#2) #3}
\newcommand{\np}[3]{Nucl.\ Phys.\ {\bf B#1} (19#2) #3}
\newcommand{\prd}[3]{Phys.\ Rev.\ {\bf D#1} (19#2) #3}
\newcommand{\prl}[3]{Phys.\ Rev.\ Lett.\ {\bf #1} (19#2) #3}
\newcommand{\prep}[3]{Phys.\ Rep.\ {\bf C#1} (19#2) #3}
\newcommand{\niam}[3]{Nucl.\ Instr.\ and Meth.\ {\bf #1} (19#2) #3}
\newcommand{\mpl}[3]{Mod.\ Phys.\ Lett.\ {\bf A#1} (19#2) #3}

\end{document}